\documentclass[10pt,conference]{IEEEtran}

\usepackage{cite}
\usepackage{graphicx}
\usepackage{amsmath}
\usepackage{amsfonts}
\usepackage{hyperref}
\usepackage{url}
\usepackage{enumitem}
\usepackage{subcaption}
\usepackage[ruled]{algorithm2e}
\SetKwComment{Comment}{/* }{ */}

\begin{document}
\title{A Deep Reinforcement Learning Framework for Optimizing Congestion Control in Data Centers}

\author{\IEEEauthorblockN{Shiva~Ketabi$^\star$ $^\dagger$,
        Hongkai~Chen$^\star$ $^\dagger$,
        Haiwei~Dong$^\star$,
        and Yashar~Ganjali$^\star$ $^\dagger$}
        \IEEEauthorblockA{$^\star$Huawei Technologies Canada, $^\dagger$University of Toronto}
}

\maketitle

\begin{abstract}
Various congestion control protocols have been designed to achieve high performance in different networking environments. Modern online learning solutions that delegate the congestion control actions to a machine cannot properly converge in the stringent time scales of data centers. We leverage multi-agent reinforcement learning to design a system for automatic and dynamic tuning of congestion control parameters at end-hosts in a data center. The system includes agents at the end-hosts to monitor and report the network and traffic states, and agents to run the reinforcement learning algorithm given the states. Based on the state of the environment, the system generates congestion control parameters that optimize network performance metrics such as throughput and latency.
As a case study, we examine BBR, an example of a prominent recently-developed congestion control protocol. Our experiments demonstrate that the proposed system has the potential to mitigate the problems of static parameters ({\em i.e.}, reduces convergence time by $2.7$x and round-trip time by $40\%$).%

\end{abstract}

\section{Introduction}

Congestion control is a crucial component in computer networking, which determines how limited shared resources (link bandwidth, buffers, etc) are divided among different network flows. This problem is extremely challenging as it should keep the total resource utilization high, while fairly serving individual flows and avoiding congestion in a distributed manner. With decades of research, and development of large number of protocols, no congestion control protocol can meet the requirements of all networks.

Network environments vary significantly in terms of latency, volatility, buffer capacity, and many other dimensions which means that a congestion control protocol should either be engineered specifically for an environment or it fails to perform optimally in that environment. Also, the dynamic behavior of traffic patterns further complicates this problem as the number of flows, traffic burstiness, and types of applications change over time, invalidating the optimal values of statically-tuned parameters.

Congestion control designers are experts in understanding the congestion control problem and defining rules for adjusting rates in a network environment. However, we cannot expect an expert to dynamically modify the rules based on the network changes. To address these problems, online learning solutions~\cite{Arura, Eagle, PCC} have been recently getting attention with the promise of generalizing the congestion control protocols and removing the need to engineer them for every environment. In practice, however, this promise is not realized. The main problem is that congestion control loops are very small, especially in data centers~\cite{DCN_RL}, making an accurate control action by an artificial agent difficult.

The high-level changes in a network environment happen orders of magnitude less frequently than congestion control loops. Therefore, we can leverage pre-knowledge congestion control protocols designed by experts, and use machines to dynamically adjust them in time scales in the order of seconds to hours. In this paper, we propose a framework for dynamically adjusting congestion control parameters in a data center. %
For the congestion control actions, happening in the order of microseconds in a data center, we use traditional protocols. In a higher-level control loop, we use Reinforcement Learning (RL) agents to determine the value of key parameters of the traditional protocols. 

More specifically, the network operator specifies the congestion control variant, the parameters that should be dynamically adjusted, and the network requirements in terms of a reward function. The end-hosts send the network and traffic state to the designed online learning module through a communication channel. The online learning module consists of a multi-agent reinforcement learning solution and periodically generates the tuned congestion control parameters.%

As a case study, we consider bottleneck bandwidth and round-trip propagation time (BBR)~\cite{BBR} as our base congestion control variant. %
BBR's performance shows promising results in Googleâ€™s environment; however, migrating BBR to other environments might degrade its performance, requiring a huge effort for readjusting the parameters. We dynamically tune two of BBR's parameters that we believe greatly affect its behavior using our implemented framework, namely RTT minimum filter window and bottleneck bandwidth maximum filter window. 

We implement our proposed framework, and test a prototype specifically for BBR. The design and implementation of our framework is generic and the evaluation can be extended to other congestion control protocols. For a network with frequent changes, while BBR can only accurately estimate the RTT for $40\%$ of the time, our RL-based framework can provide a correct estimation for more than $60\%$ of the testing time. Moreover, for a scenario with multiple flows, it reduces the convergence time of BBR by $2.7$x and the peak RTT by $40\%$.%

\section{Related Work}
In this section, we first discuss the challenges of traditional congestion control protocols. Next, we introduce some of the seminal works on using RL for congestion control and compare our proposed solution with them.

\subsection{Traditional Congestion Control}
Today's networks extensively vary in terms of different properties including link capacities, RTTs, topology structure, loss-rate, frequency of changes, and buffer sizes~\cite{PCC,OpenTcp}. As a result, many congestion control protocols have been specifically designed for a limited range of network properties. Protocols that perfectly addressed the congestion control problem in the 80s and 90s, soon became inefficient for the networks emerging a few years later with high bandwidth-delay product. As a response to this change, some new protocols with more aggressive congestion window increase functions were introduced (e.g. BIC~\cite{bic} and CUBIC~\cite{cubic}). Others tried to manually adjust the parameters of the existing protocols based on the new network requirements. In the original NewReno~\cite{newreno}, for example, the initial congestion control window was set to $1$ segment, that was later changed to $2$, $4$, and $10$ segments in order to increase the utilization as the network bandwidth-delay product has been increasing.

Moreover, the state of a single network dynamically changes over time due to changes in traffic patterns, type of applications, link failures, path changes, {\em etc.} However, the parameters of congestion control protocols are statically adjusted and used during the network operation. As an example, in BBR, RTT minimum filter window and bottleneck bandwidth maximum filter window are set to $10$ seconds, and $8$ RTTs, respectively. The optimal configuration of these parameters depend on the dynamic state of the network and the traffic patterns. In the absence of a dynamic adjustment to congestion control parameters, inefficient behavior will be observed in a dynamically changing network.

\subsection{Reinforcement Learning for Congestion Control}
Reinforcement learning solutions have been proposed for decision making in dynamic, complex, and unstructured environments ({\em e.g.},~\cite{gu2017deep}). %
A new direction to deal with the dynamic changes in the network is to let RL solutions determine the congestion control actions instead of relying on human-designed congestion control protocols~\cite{Arura, Eagle, lei2020congestion}.

These RL solutions are required to provide congestion control actions in RTT time scales. The hope of these solutions is to avoid hard-wiring network signals and control actions, and to automatically learn the optimal control actions by interacting with the network. %
However, these solutions do not generalize to unseen network conditions and converge much slower than the network dynamics~\cite{classicMeetsModern}. The reason is the high frequency and range of changes that make the time and the data limited for the algorithm to learn about the huge state space and complex dynamics of a computer network.
Recently, it is suggested to use RL techniques for more coarse-grained decisions as opposed to directly using them in congestion control loops. Orca~\cite{classicMeetsModern}, for example, uses a traditional protocol as its base congestion control mechanism and plugs an RL agent for adjusting the rates in time scales much larger than the congestion control loop. ACC~\cite{ACC}, another hybrid approach, adjusts the Explicit Congestion Notification (ECN) threshold values for ECN-based protocols using multi-agent RL techniques.

Unlike Orca, we do not directly intervene in the congestion control loop and do not touch the congestion control window. We, however, only adjust the parameters of the congestion control protocols based on the dynamics of the environment and network conditions. Comparing our solution with ACC, our design focuses on the congestion control parameters at the end-hosts to avoid changes to the core switches in a data center resulting in a more deployable solution. Our framework is not limited to any specific parameter (such as ECN threshold in ACC), and can be generalized to any host-side parameter of interest.

\begin{figure*}[t]
\centering
\includegraphics[width=\textwidth]{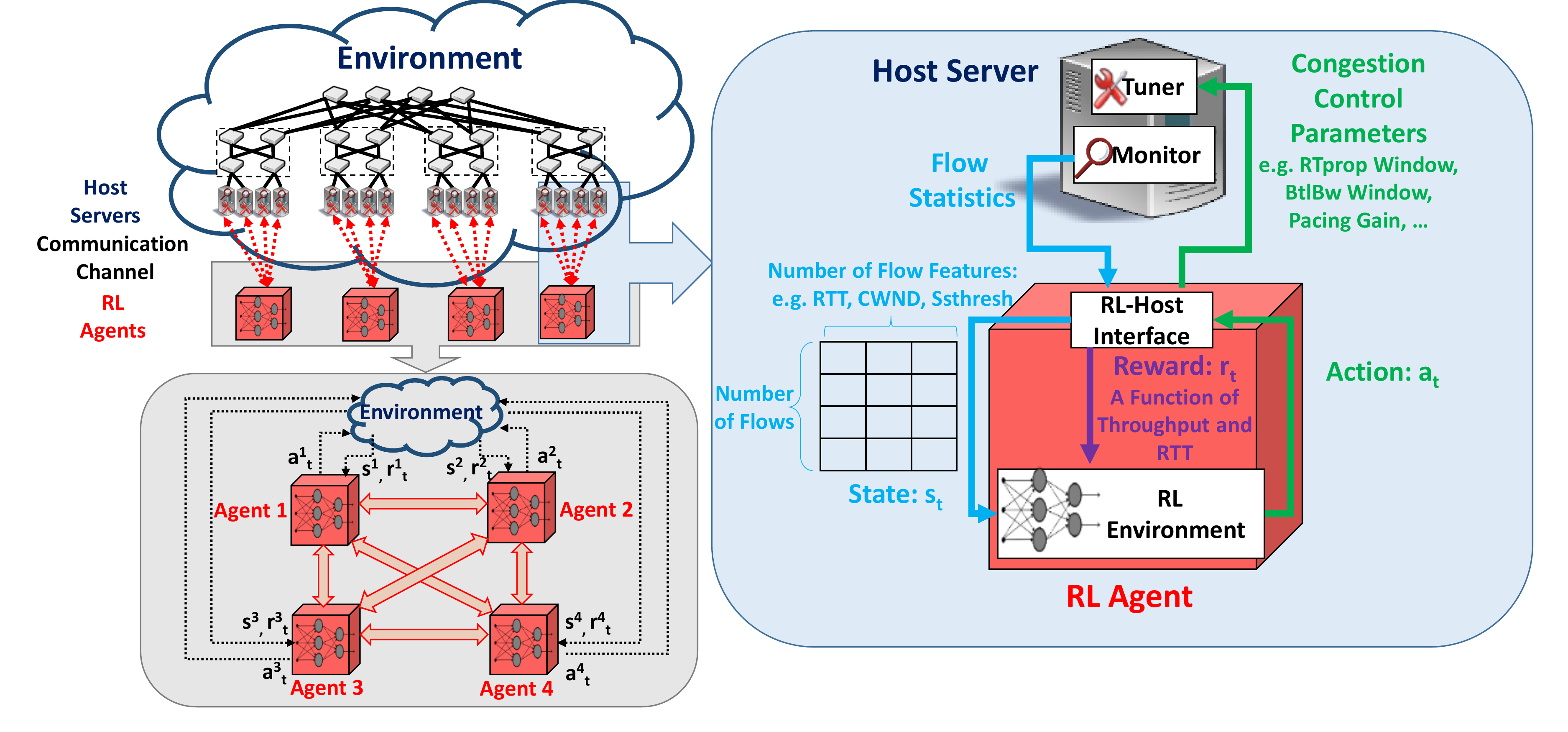}
\caption{The system architecture and components. On the top left, an example of a data center topology is illustrated as our environment. In the proposed architecture, host servers are added to the end-hosts and are connected to a number of RL agents via a communication channel. On the right side, we focus on the communication between the components of the host servers with the RL agents, and the details of the RL formulation. On the bottom left, the figure zooms in on the network between the multiple RL agents.}
\label{fig:system}
\end{figure*}

\section{System Architecture} 
\label{sec:sys}
In this section, the design principles for our framework as well as the architecture and component details are discussed.

\subsection{Design Principals and Considerations}
{\noindent\bf Local vs. Global Optimization.} 
Typically, online learning protocols~\cite{Arura, Eagle, PCC, classicMeetsModern} focus on networks with long delays, such as core networks or even Internet scale, to give the RL agents enough processing time. The problem, however, is that it is almost impossible to plan for a global optimization in a wide-area network such as the Internet. The reward function in these protocols is defined per flow and there is no explicit communication about the state of different flows. %

On the contrary, data center networks are controlled by a single authority, making it much easier to deploy the RL agents throughout the network, to optimize toward a common global goal, and to spread the information between different agents. The challenge in a data center is the stringent time scales and the extremely dynamic nature of a data center as previously mentioned. Our solution addresses these challenges by relying on human-designed congestion control protocols, and dynamically adjusting the parameters of these protocols in larger time scales compared to the congestion control cycle.

{\noindent\bf Multi-Agent Cooperative Learning.}
Data centers can host thousands of switches and servers and the required state space of an RL solution grows proportionally with the scale of the network. To design a scalable system, we employ a fully cooperative notion of multi-agent RL in which all agents operate in the same environment and work for achieving a long-term common goal~\cite{zhang2021multi}. The number of RL agents depends on the scale of the data center and is not fixed. 

As shown in the bottom left of Figure~\ref{fig:system}, each agent receives a partial observation and reward from the environment and some information from other connected agents, a.k.a, its neighbors. This information includes shared sensation ({\em i.e.}, flow stats), shared episodes (trajectories of [state, action, reward]), and shared learned policies. Each agent independently optimizes based on its partially observed information and generates a partial action that is the value of congestion control parameters for a subset of the flows controlled by the agent. Then, after a time interval the information are shared among different agents. During the interval, we do not require the agents to converge to the same action. In our fully cooperative learning setting, all RL agents share the same reward function meanwhile optimizing the team-average reward. The difference in value estimation between an RL agent and its neighbors is included as a penalty term in its value function update.%

{\noindent\bf Robustness in Deployment.} When utilizing reinforcement learning in deployment, robustness is very important and we have to remove as many unreliable statistics and outliers as possible. For this purpose, the first method is applying a low pass filter to remove high-frequency perturbation ({\em e.g.}, latency jitters). The cut-off frequency can be easily predefined according to the network environment. The second method is using regression to smooth the network measurements (such as $\frac{d(RTT)}{dT}$ which is sensitive to RTT changes) where moving window average is highly efficient. Another helpful technique is verifying the rationality of the tuple [network measurement, action, and reward], by setting a number of fuzzy rules before feeding it to the RL algorithms.%

\subsection{System Design}
The proposed system consists of two main components: host-servers and RL agents. We enable a communication channel between the host-servers and the RL agents. The architecture and components of the system, demonstrated in Figure~\ref{fig:system}, are explained in details below.

{\noindent\bf Host Servers.} We add agents at the end-hosts for monitoring flow stats and enforcing the tuned parameters. Each host server includes the two following components:
\begin{itemize}[leftmargin=*]
\item Monitor: The monitor is responsible for monitoring flow stats including throughput, delay, and loss. It samples the flow stats every $T_1$ seconds ($T_1$ is in the order of milliseconds), and updates the local state of the host. The value of $T_1$ introduces a trade-off between accuracy and overhead of the sampling.%
\item Tuner: The tuner receives the tuned congestion control parameters every $T_2$ seconds from the associated RL agent. $T_2$ is in the order of seconds to hours (Figure 1 in~\cite{OpenTcp} suggests that the high-level traffic pattern changes happen in the order of hours). The tuner includes a channel with the kernel congestion control module to send the new values from the user space to the kernel space.%
\end{itemize}

{\noindent\bf RL Agents.} We dedicate servers for running the RL algorithm and training the RL model in an online manner. We can have one or multiple servers based on the scale of the data center. In addition, in a software-defined network, we can leverage control plane resources for this purpose.

\begin{itemize}[leftmargin=*]
\item RL-Host Interface: This interface is responsible for creating a communication channel between the host servers and the RL agents. It wraps up the parameters into an RPC message to the host servers and waits for receiving the stats in response in $T_2$ seconds. It also translates the input/output to/from the RL agents to the proper format. More specifically, it receives flow stats from different associated hosts and stores them in a 2D array with rows representing flows and columns representing flow stat features. This constructs the RL state at each step. It also calculates the RL reward and sends both the state and the reward to the RL environment model. The optimal actions received at each step are converted to congestion control parameters and transferred to the associated hosts.

\item RL Environment Model: The RL environment model is the core of the RL agent that iteratively generates the optimal actions given the state and the reward at each time using an RL training algorithm. %
 It stores the neural network model which can be a Q-network, a policy network, an actor network, or a critic network based on the applied RL training algorithm.
\end{itemize}

\section{Problem Formulation}
\label{sec:problem}

The problem is modeled as an environment and one or more agents interacting with the environment by observing the state, taking actions, and receiving rewards. At each time $t$, the {\em state} of the environment is represented by $s_t$, the taken {\em action} by $a_t$, and the received {\em reward} for taking action $a_t$ in state $s_t$ by $r_t$. The environment model specifies the {\em transition function} which receives a state $s_t$ and an action $a_t$, and outputs the probability of the next state being $s_{t+1}$. The transition function is assumed to have the Markov property.

\subsection{Optimization Problem} 
At each time, each agent is responsible for suggesting the best action in each state, with the goal of maximizing the sum of future discounted reward ~\cite{sutton2018reinforcement}:
\begin{equation}
\max_{\pi_{\theta}}\sum_{t}^{\textit{Total Time}} \gamma^{t}r_t
\label{eq1}
\end{equation}
Here, $\gamma$ is the discount factor which prioritizes the rewards received closer to the current time. To find the best action at each time, the agent optimizes its {\em policy} $\pi_{\theta}$ in which parameter $\theta$ represents the weights of a deep neural network.%

In what follows, we will define the RL formulation of our system including the state space, the action space, and the reward function.

{\noindent\bf State Space.} The state at each time $t$, denoted by $s_t$, shows the flow statistics including the measurements of individual flows from the network. It is visualized as a table in Figure~\ref{fig:system}. Each row is dedicated to a single flow and each column shows flow features such as RTT, delivery rate, and congestion window size. If the number of flows is less than the maximum table capacity, the extra space will be zero-padded. If the number of flows exceeds the table capacity, a sample of the flows equal to the table capacity will be selected. Sampling flows (with a fixed size) for the state space of each RL agent and extending the number of RL agents for a large scale network ensure the scalability of the model.

For our case study of BBR explained in Section~\ref{sec:BBR}, in addition to the aforementioned features, we take BBR variables into account for the state space including BtlBw, RTprop, pacing gain, and CWND gain.

{\noindent\bf Action Space.} The action at each state, $a_t$, is defined as an array of all the congestion control parameters that should dynamically be adjusted $a_t = [a_t[1], \dots, a_t[M]]$ where $M$ shows the total number of adjustable parameters.

In our experiments, we focus on two parameters ($M=2$) of BBR corresponding to the size of the RTprop window ($a_t[1]$) and BtlBw window ($a_t[2]$). These parameters represent the window sizes for keeping the minimum RTprop and the maximum BtlBw estimations.%

{\noindent\bf Reward Function.} The objective of any congestion control protocol is to achieve high throughput and low latency. Therefore, the reward function, {\em i.e.}, $r_t$, should be a function of throughput, latency, and other metrics of interest based on the congestion control variant and the network environment. For the reward function, we can combine the proportional factor (P), the derivative (D), and/or the integral (I) over time of the metrics~\cite{PID}. The P term helps to balance the weight of different metrics in the reward function, {\em e.g.}, $K_P \cdot \frac{throughput}{latency}$. The D term can help to reduce overshooting or oscillating, {\em e.g.}, $K_D \cdot \frac{d(RTT)}{dT}$ \cite{PCC}. The I term is used to reduce the accumulative estimation error, {\em e.g.}, $K_I \cdot \int ({latency}_{estimate}-latency) dt$.

Particularly, since BBR's goal is to estimate the values of BtlBw and RTprop as accurately as possible, the reward function is made up of the measured link utilization and the difference between the actual RTT and the estimated RTprop during the interval ($T_2$): $r_t = \alpha \times Throughput + (1-\alpha) \times 1/(1+\exp(|{latency}_{estimate}-latency)|)$. We apply a negative sigmoid function to the second part to significantly increase the gap between small and large values. There is also a constant coefficient for each part to adjust the weights.

\subsection{Deep Reinforcement Learning Controller}
We first explain the domain knowledge that led to the choice of the RL controller algorithm. Timing is extremely critical in computer networks especially data centers. RL algorithms with slow convergence, even with the best performance, are not suggested for this domain. As the environment is highly sensitive and real-time, the stability of the RL algorithm is another important decision factor. In addition, a data center can be as large as thousands of devices, thus for a scalable solution, we require an RL algorithm with high performance but low resource requirements and simple implementation.

To this end, we use Proximal Policy Optimization (PPO)~\cite{PPO} for the RL controller algorithm. PPO has shown fast convergence with relatively low sample complexity, and high performance results. Compared to other algorithms with similar performance, PPO is much simpler to implement and deploy, is less sensitive to hyper-parameters, requires lower number of samples and processing steps to converge, and does not need any memory for replay buffers. In addition, PPO's optimization function is designed such that the policy does not deviate much after each update, ensuring the stability of the algorithm. All these properties, makes PPO the best match for our domain.

Let us review how the objective function is constructed for PPO. Along with the advantage estimations, it includes a probability ratio: $R_t(\theta)=\frac{\pi_{\theta}(a_t|s_t)}{\pi_{\theta_{old}}(a_t|s_t)}$ in which $\pi_{\theta_{old}}$ is the old policy and $\pi_{\theta}$ is the policy after the update. To avoid large policy updates, PPO clips the objective within a range of the advantage estimation. The objective function is defined as follows:

\begin{align}
\label{eq2}
L^1_t = \mathbb{E}[\min(&R_t(\theta)\hat{A}_t, \text{clip}(R_t(\theta),1-\epsilon,1+\epsilon)\hat{A}_t)]
\end{align}

$\hat{A}_t$ is the estimated advantage at time t and $\epsilon$ is a hyperparameter controlling the clip range. The advantage function is the difference of the expected value of a state given an action and the expected value of the same state over all the possible actions. For estimating the advantage function, PPO uses a neural network model (critic) to train a value-state function, which is a function over states showing how much total reward is expected to be achieved starting at a given state.

The Generalized Advantage Estimation (GAE) in Algorithm~\ref{alg:GAE} calculates advantage estimations. It receives a set of trajectories, and starting from the reverse direction of time, it calculates the advantages per each time using the value estimations of the critic network. Here, $\lambda$ is a smoothing factor for reducing the variance.

To train the critic model, a squared-error loss function should be included in the optimization function:
\begin{align}
\label{eq3}
 L^2_t = (V_{\theta}(s_t)-V_t^{target})^2
\end{align}

Combining the actor loss, $L^1_t$ in Equation~\ref{eq2}, with the critic loss, $L^2_t$ in Equation~\ref{eq3}, and an entropy term for exploration, {\em i.e.},:

\begin{align}
\label{eqS}
S = \beta_{entropy} \mathbb{E} [- \pi_{\theta}(a_t|s_t)\log{\pi_{\theta}(a_t|s_t)}]
\end{align}
we can achieve the following surrogate objective function for Equation~\ref{eq1} in which $c_1$ and $c_2$ are coefficients:

\begin{align}
\label{eq4}
\max_{\theta} \mathbb{E}[L^1_t - c_1 L^2_t + c_2 S]
\end{align}

The PPO algorithm~\cite{PPO} is presented in Algorithm~\ref{alg:PPO}, indicating how the policy is updated according to the objective function. During each iteration, N actors independently collect data of the 
observed trajectories for $T$ timesteps that is sequence of states, actions, and rewards. Given the trajectories, the GAE algorithm computes the advantage estimations for each time required for the policy update. Finally, using the collected data and the estimated values, it finds a new value for $\theta$, maximizing the surrogate function in Equation~\ref{eq4} using Stochastic Gradient Dissent (SGD) (or similar methods).

\begin{algorithm}[t]
\SetAlgoLined
\SetKwInOut{Input}{Input}
\SetKwInOut{Output}{Output}
\Input{$\{(s_t, a_t, r_t)|0\leq t\leq T-1\}$}
\Output{$\{\hat{A_1},\dots,\hat{A_T}\}$}

$\hat{A_T} \gets 0$\;
\For{$t\leftarrow T-1$ \KwTo $1$}{
Estimate $V(s_t)$ with critic network\;
$\delta_t \gets r_t+\gamma V(s_{t+1})-V(s_t)$\;
$\hat{A_t} \gets \delta_t+\gamma \lambda \hat{A}_{t+1}$\;
}
\caption{Generalized Advantage Estimation}
\label{alg:GAE}
\end{algorithm}

\begin{algorithm}
\SetAlgoLined
\SetKwInOut{Input}{Input}
\SetKwInOut{Output}{Output}
\Input{The Environment, Initial $\theta$}
\Output{$\theta$}

\For{$iteration \leftarrow 1$ \KwTo $\dots$}{
	\For{$actor \leftarrow 1$ \KwTo $N$}{
		Run policy $\pi_{\theta_{old}}$ in environment for $T$ timesteps\;
		Collect states, actions, and rewards\;
		Compute $\hat{A_1},\dots,\hat{A_T}$ using Algorithm~\ref{alg:GAE}\;
	}
	Update the policy by maximizing Equation~\ref{eq4}\;
	$\theta_{old} \gets \theta$\;
	$\theta=\underset{\theta}{\arg\max}\frac{1}{NT}\underset{\tau}{\sum} \sum_{t=1}^{T} (L^1_t - c_1 L^2_t + c_2 S)$\;
}

\caption{Proximal Policy Optimization}
\label{alg:PPO}
\end{algorithm}

\section{System Prototyping and Evaluation}
\label{sec:BBR}
In this section, we first review BBR which is the base congestion control protocol in our evaluation. A simple experiment is provided to show the shortcomings of static parameter settings in BBR and to motivate for our RL-based framework. Then, we explain our proof-of-concept implementation. We dynamically tune two of BBR parameters using our prototype and present experimental results for this case study. The same implementation can be extended to tune and evaluate other congestion control protocols.%

\begin{figure*}[t]
	\centering
	\begin{subfigure}{.48\textwidth}
		\includegraphics[width=\textwidth]{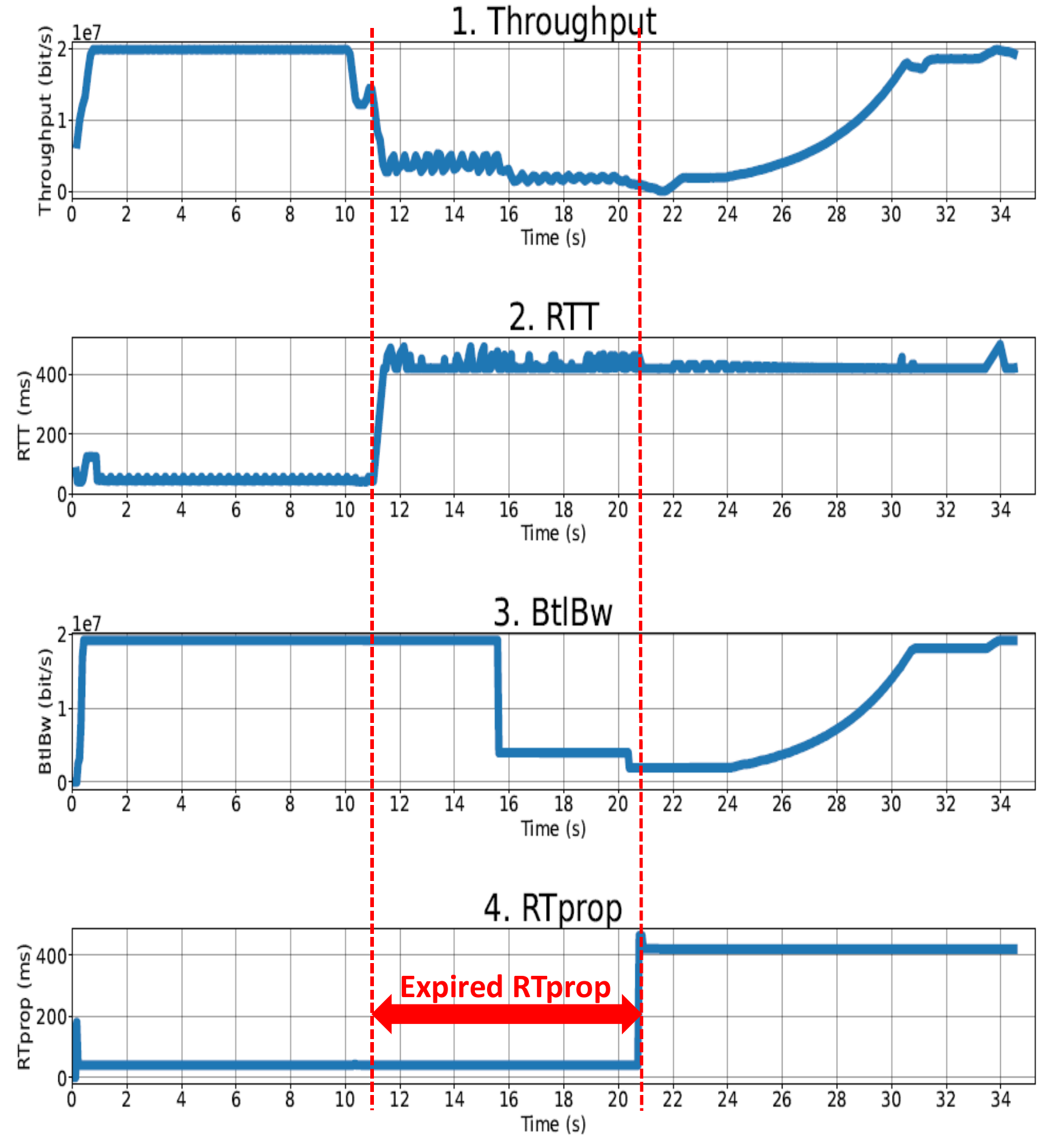}
		\caption{The effect of a large static value for RTprop window on RTprop estimation and throughput.}
		\label{fig:BBR_problem_Rtprop}
	\end{subfigure}
	\begin{subfigure}{.48\textwidth}
		\includegraphics[width=\textwidth]{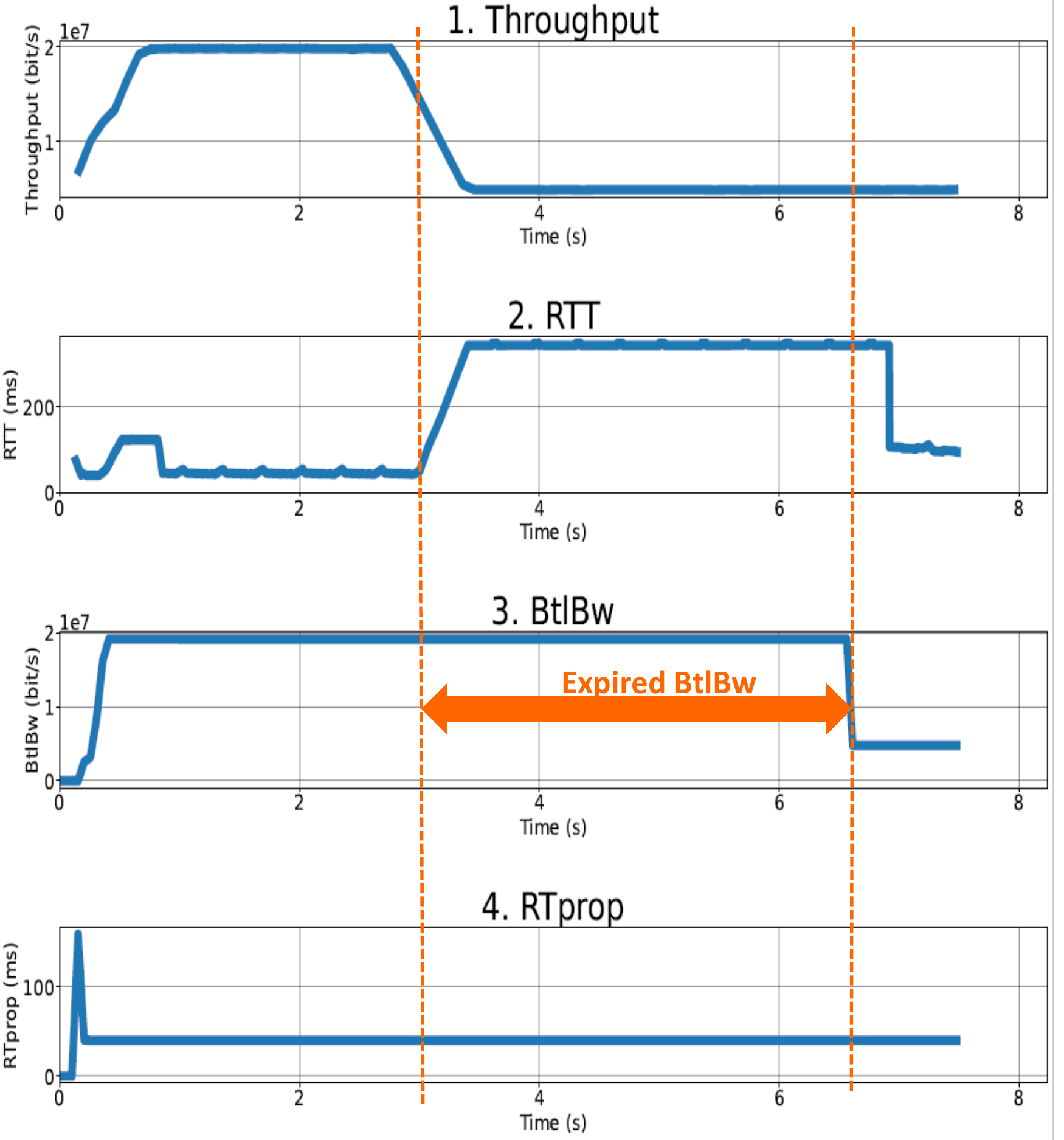}
		\caption{The effect of a large static value for BtlBw window on BtlBw estimation, throughput, and RTT.}
		\label{fig:BBR_problem_BtlBw}
	\end{subfigure}
	\caption{BBR case study.}
\label{fig:BBR_problem}
\end{figure*}

\subsection{Case Study: BBR}
A wide range of traditional congestion control protocols infer loss as a congestion signal and define their dynamics based on observing a packet loss. The drawback is that a packet loss might happen due to different reasons one of which is congestion. Furthermore, reacting based on a packet loss sets the operating point of the system at the full capacity of the buffers, resulting in large packet latency. BBR has been proposed to directly address this issue by operating at a point with minimum latency and maximum utilization. For this purpose, BBR keeps an estimation of the minimum Round-Trip Propagation time (RTprop) and the maximum Bottleneck Bandwidth (BtlBw) over two windows with fixed sizes, and keeps the inflight traffic proportional to BtlBw times RTprop.

If RTprop is increased or BtlBw is decreased, the estimation will not change for the rest of the associated window, thus becomes invalid. To illustrate the effects of this problem, we emulate a flow with a bottleneck link capacity of $20$Mbps and an initial RTT of $40$ms using the framework developed in~\cite{BBRFramework}. After $11$s, we change the RTT to $400$ms which causes the RTprop estimation to get expired. As shown in Figure~\ref{fig:BBR_problem_Rtprop}, the RTprop stores the invalid minimum for RTT for around $9-10$s and mistakenly thinks the bottleneck bandwidth is reduced, leading to a drop in throughput for around $20$s. Furthermore, Figure~\ref{fig:BBR_problem_BtlBw} presents another emulated experiment in which we reduce BtlBw from $20$Mbps to $5$Mbps at time $3$s. As the figure depicts, the BtlBw estimation is invalid during $3$s-$7$s interval when the change leads to a buffer build-up and a huge increase in RTT. This experiment shows that BBR is not operating at the desired operating point since the buffer occupancy is high at least for a few seconds.

The static parameter setting of BBR has led to promising results in Googleâ€™s wide area network; however, migrating BBR, as is, to other environments outside Google may cause issues. Specifically, it might be challenging to migrate BBR to a data center environment. In a data center, the two aforementioned scenarios are very frequent: changes in link capacities, {\em e.g.}, due to changes in traffic patterns, and changes in RTT, {\em e.g.}, due to changes in flow paths. Also, a few seconds of inefficient behavior in a data center is more troubling compared to a wide area network as extremely short flow/task completion times are required.

\subsection{Proof-of-Concept}
\label{sec:prototype}
We have implemented a system prototype based on the aforementioned architecture. This prototype can be integrated with any congestion control protocol that exits in Linux kernel. For our evaluation, we focus on tuning of BBR. The prototype is built of different components as introduced in Section~\ref{sec:sys} and Section~\ref{sec:problem}. For the monitor located at the host servers, we leverage Linux ss command which is a tool that provides network statistics. The tuner component is written in C and communicates with congestion control kernel models using ioctl system call to adjust the parameters on demand.  

The implementation of the interface between the host servers and the RL agents, and the RL environment model are in Python. For communication between the host server and RL agent we use ZeroMQ asynchronous messaging library. Also, we use the OpenAI Gym common interface for implementing the RL environment model in combination with the Stable Baselines as the library for running the RL algorithms.

For the network emulation, we use a publicly available framework for reproducible TCP measurements in Mininet emulation environment~\cite{BBRFramework}. It provides a flexible mechanism to test different network settings and create measurement plots.

\begin{figure*}[t]
	\centering
	\begin{subfigure}{.47\textwidth}
		\includegraphics[width=\textwidth]{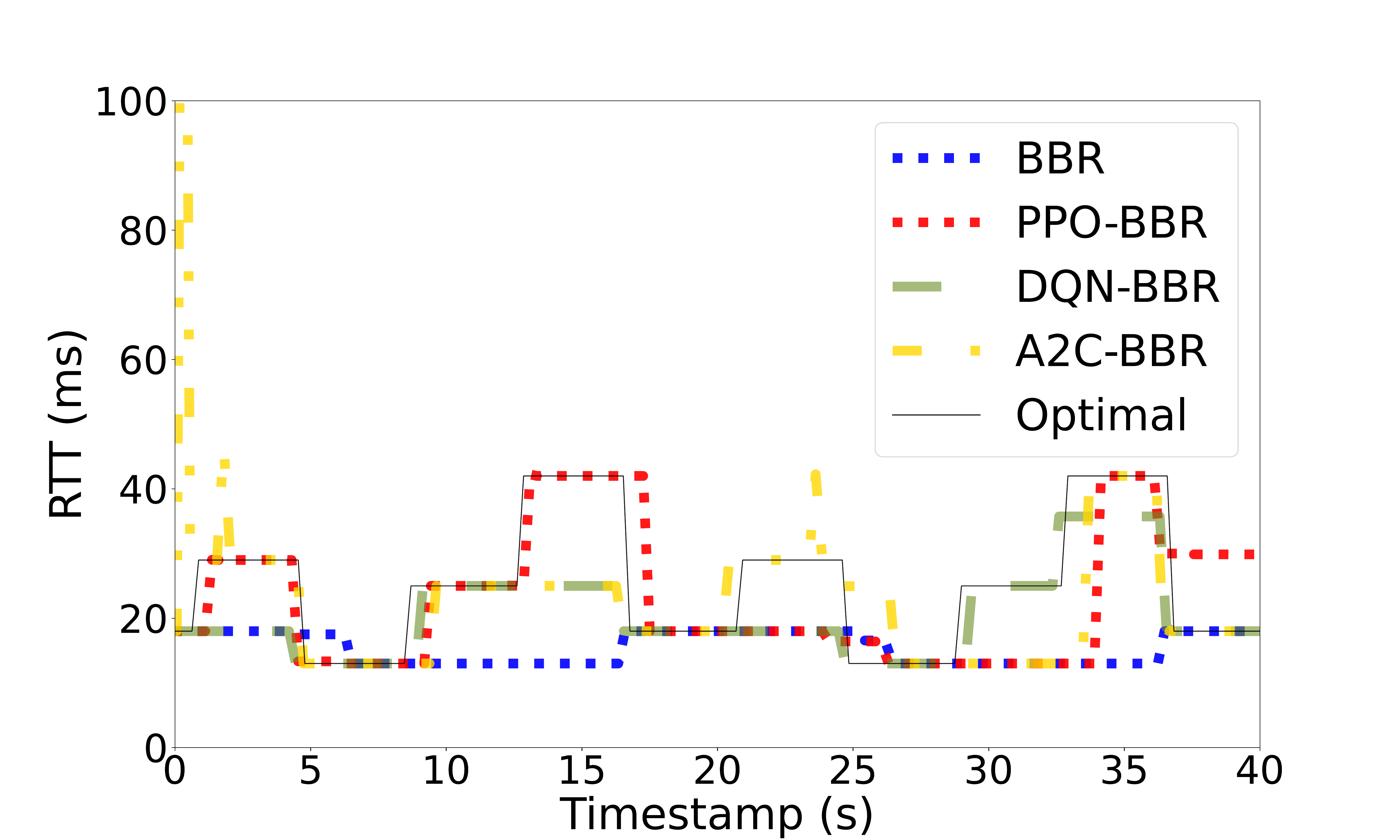}
		\caption{ }
		\label{fig:single_flow:a}
	\end{subfigure}
	\begin{subfigure}{.47\textwidth}
		\includegraphics[width=\textwidth]{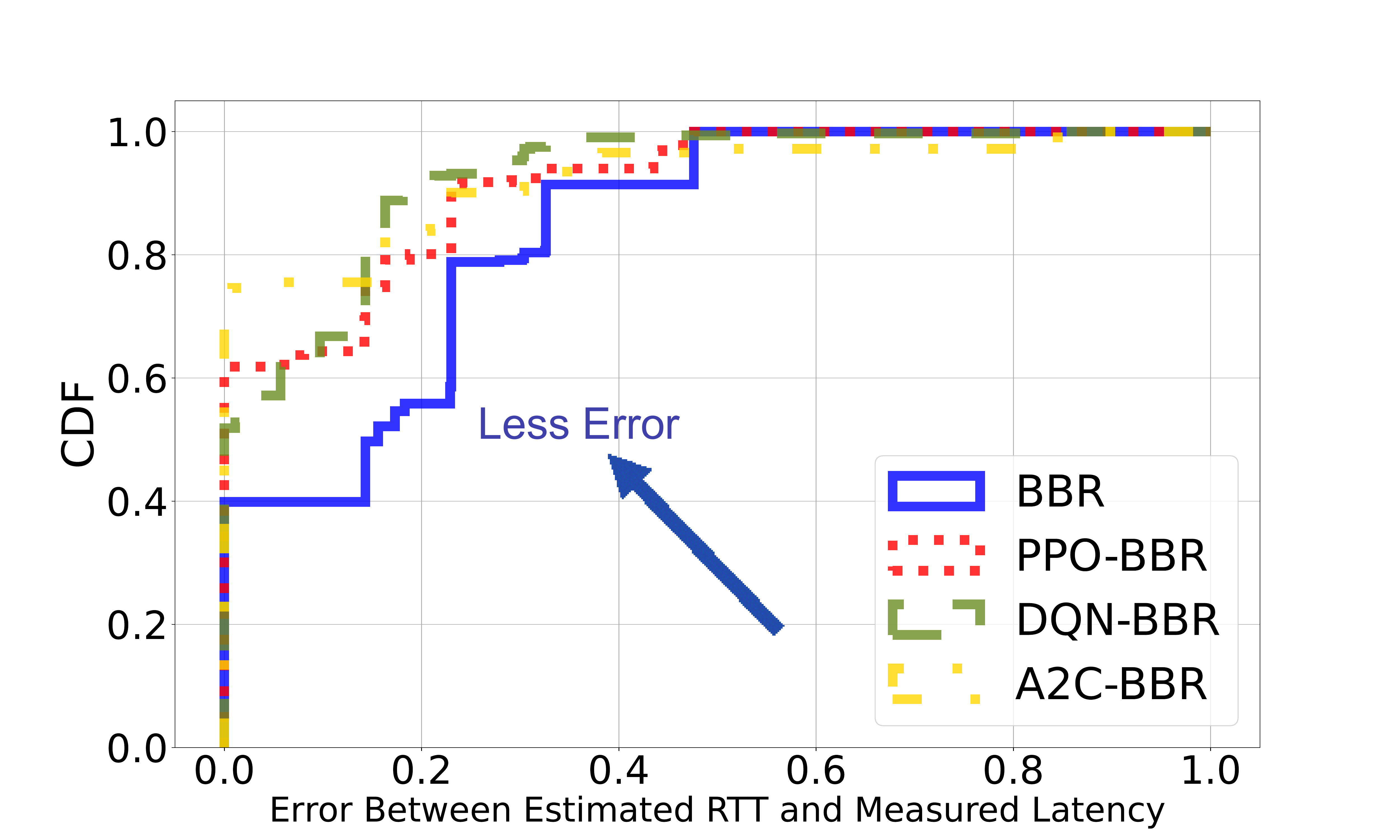}
		\caption{ }
		\label{fig:single_flow:b}
	\end{subfigure}
	\begin{subfigure}{.47\textwidth}
		\includegraphics[width=\textwidth]{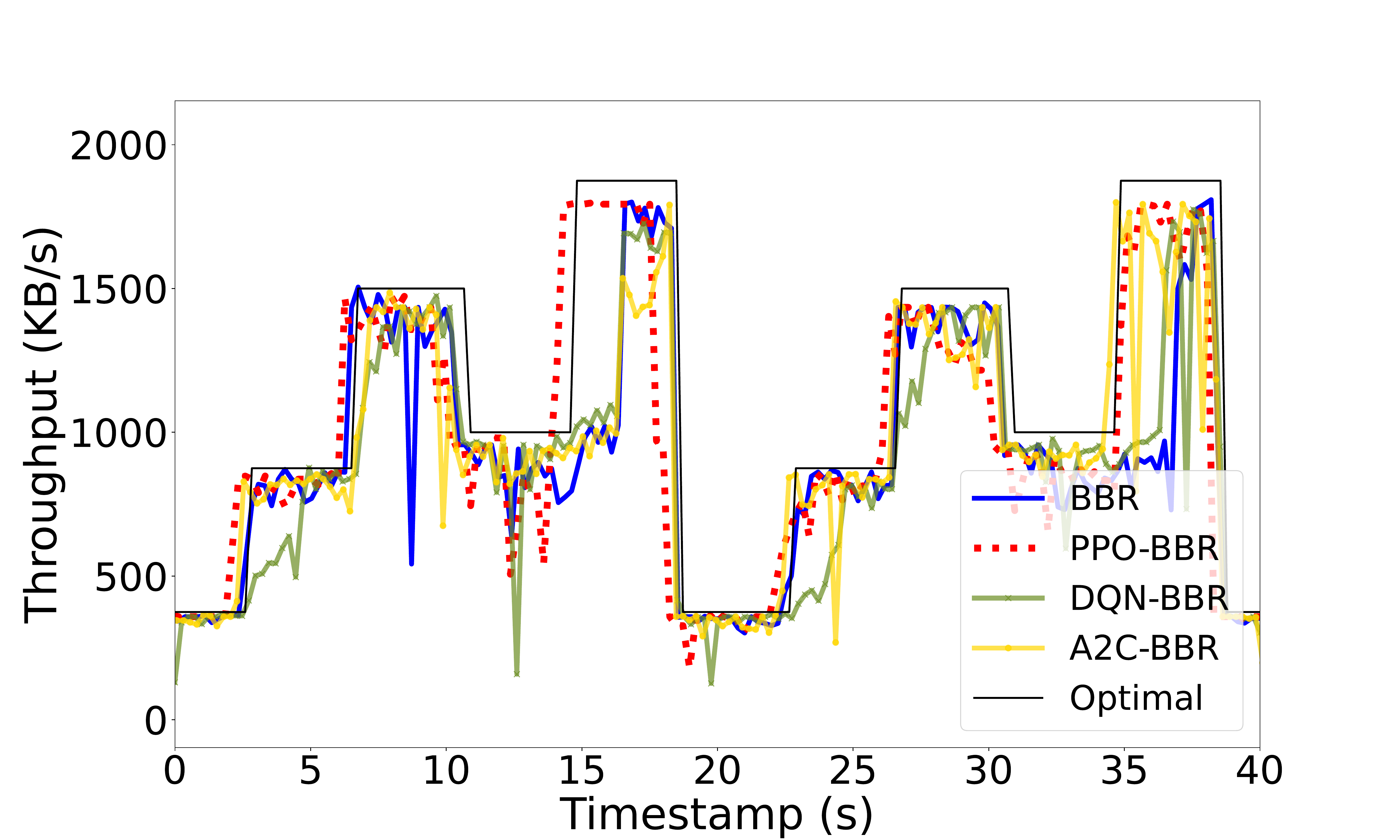}
		\caption{ }
		\label{fig:single_flow:c}
	\end{subfigure}
	\begin{subfigure}{.47\textwidth}
		\includegraphics[width=\textwidth]{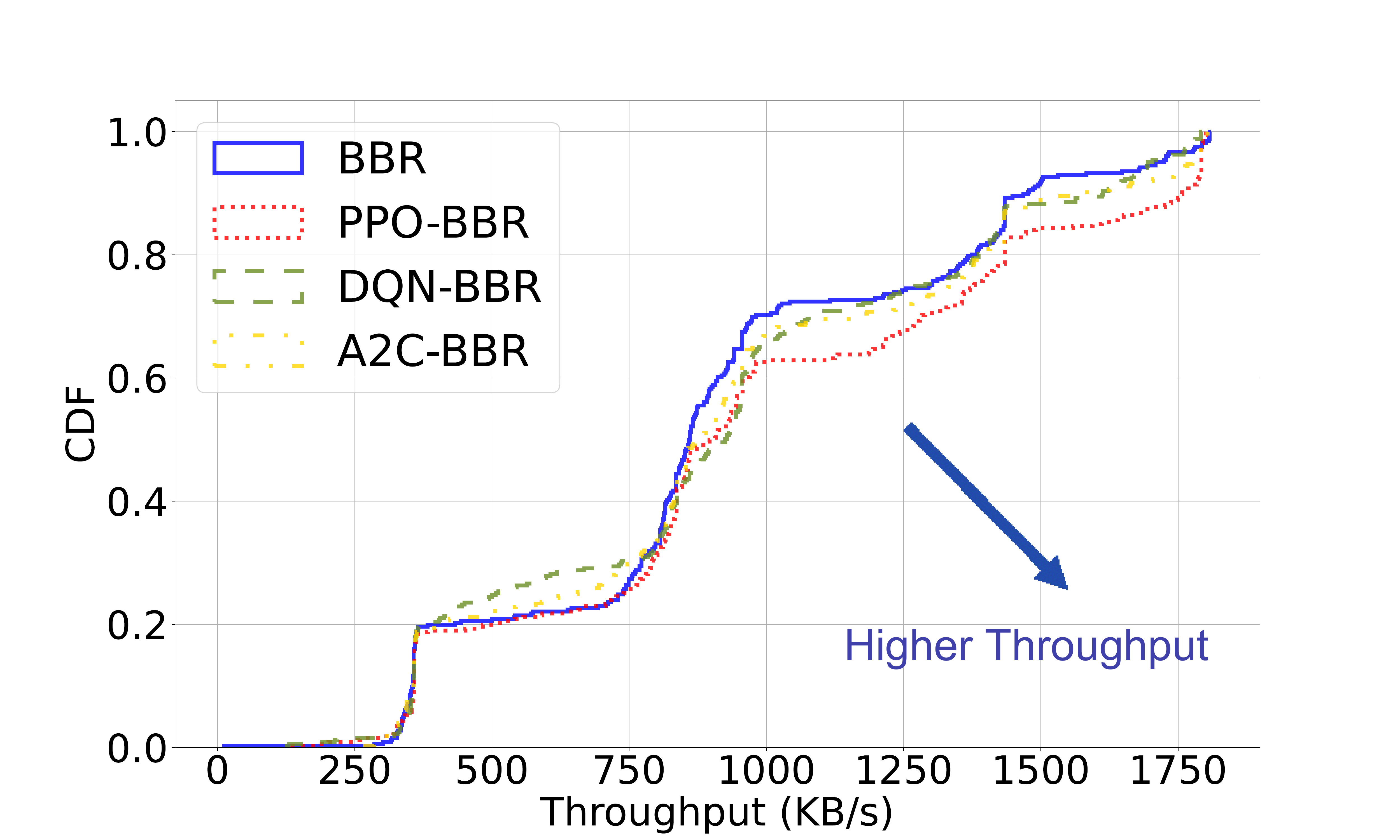}
		\caption{ }
		\label{fig:single_flow:d}
	\end{subfigure}
	\caption{Comparison of the vanilla BBR, PPO-BBR, DQN-BBR, and A2C-BBR in terms of (a) estimated RTT, (b) CDF of the error between the estimated RTT and latency, (c) throughput, and (d) CDF of the throughput.}
\label{fig:single_flow}
\end{figure*}

\subsection{Experimental Results}
\label{sec:experiments}
We evaluate our prototype for dynamically adjusting BBR's windows for maximum BtlBw and minimum RTprop estimations. To train the RL agent, we generate a series of random events where the available bandwidth varies from $1$ to $10$Mbps, latency varies from $10$ms to $50$ms, and flows joining and leaving the network. The random intervals of these events are uniformly distributed between $1$ second and $50$ seconds. We train the agent under such workload for $1.5$ hours with a learning rate of $3\times 10^{-3}$ and evaluate it online with a learning rate of $10^{-4}$.

{\noindent\bf Comparison of PPO-BBR with other BBR Solutions.} To understand the effects of different RL algorithms, we first focus on the evaluation of a single flow. During each test epoch, the latency and bandwidth change approximately every $4$ seconds. We compare our PPO-BBR with vanilla BBR and other RL-based algorithms in terms of throughput and accuracy of minimum RTT estimation in Figure~\ref{fig:single_flow}.

As shown in Figure~\ref{fig:single_flow:a}, the vanilla BBR often underestimates RTT due to its ignorance of an increase in the physical latency of the network. We argue that frequent latency changes are hardly perceived by BBR since it has fixed parameters designed for its own testbed. However, all RL-based approaches can at least detect an increase or decrease in latency once during our test. PPO-BBR can detect these latency changes more frequently without having significant overestimations compared to other RL-based approaches.

In addition, we compute the squared error between the estimated and the actual RTT, and show its cumulative distribution function (CDF) in Figure~\ref{fig:single_flow:b}. While the vanilla BBR can only accurately estimate the current RTT for $40\%$ of the testing time, PPO-BBR, DQN-BBR, and A2C-BBR
can provide a correct estimation for more than $60\%$, $50\%$, and $75\%$ of the testing time, respectively. This observation shows RL-based BBRs improve the accuracy of latency estimation.

As the throughput is shown in Figure~\ref{fig:single_flow:c}, BBR with static parameters can utilize the increased bandwidth but with a delay. The vanilla BBR never achieves high throughput in the first $2$ seconds after an increase in bandwidth whereas PPO-BBR and A2C-BBR can almost immediately acquire the bandwidth. Due to the BBR's nature, a more frequent probe cycle can lead to more drops in the throughput. However, PPO-BBR more quickly reacts to throughput changes with the same number of drops in throughput.

To obtain a more comprehensive view, we plot the CDF of throughput in Figure~\ref{fig:single_flow:d}. The vanilla BBR can reasonably utilize the bandwidth, but its throughput during the test is always worse than PPO-BBR and worse than all RL-based BBRs for throughput values higher than $800KB/s$. %
RL-based approaches show a faster response to bandwidth changes since they use observations and reward values from the network for adjusting the BtlBw window size. Our experiment demonstrates that PPO-BBR matches the performance of the vanilla BBR for small bandwidth settings and exceeds BBR's throughput in large bandwidth settings.

\begin{figure*}[t]
	\centering
	\begin{subfigure}{0.9\textwidth}
		\includegraphics[width=\textwidth]{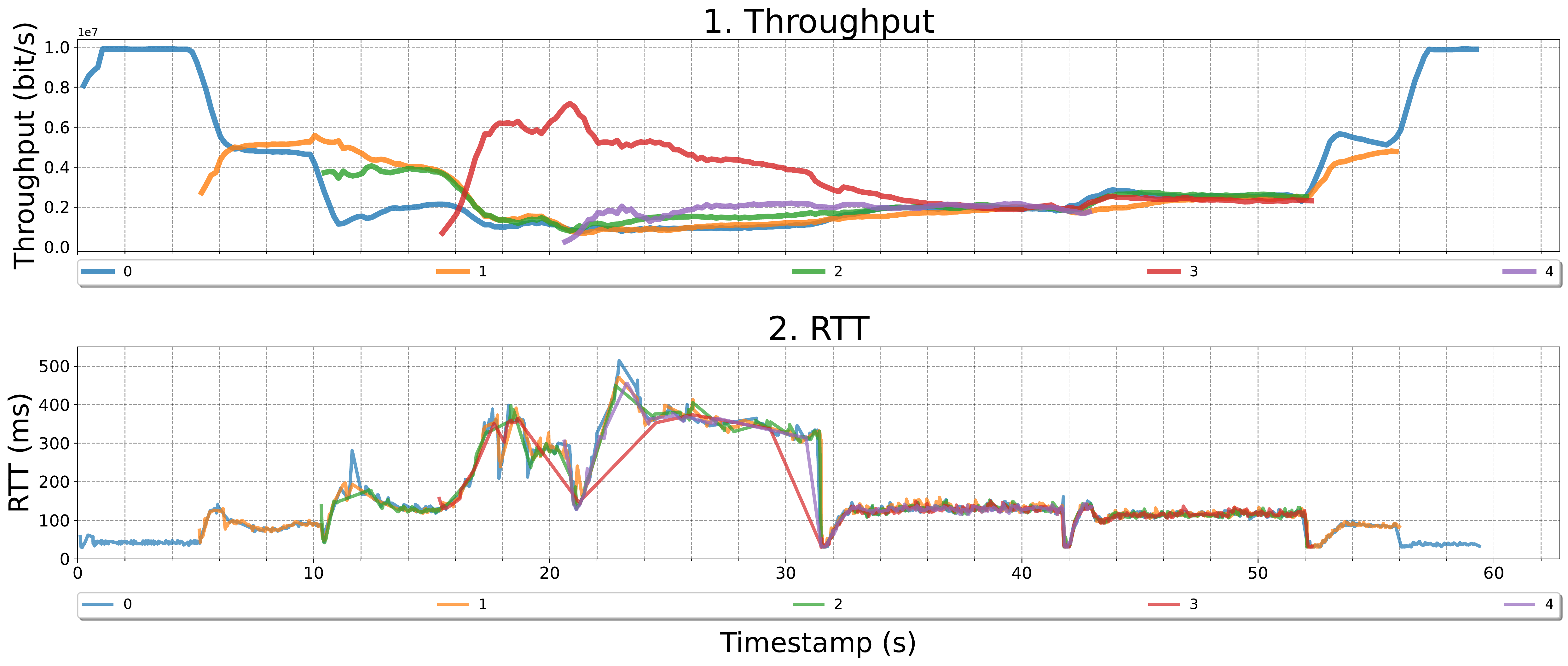}
		\caption{Vanilla BBR}
	\end{subfigure}
	\begin{subfigure}{0.9\textwidth}
		\includegraphics[width=\textwidth]{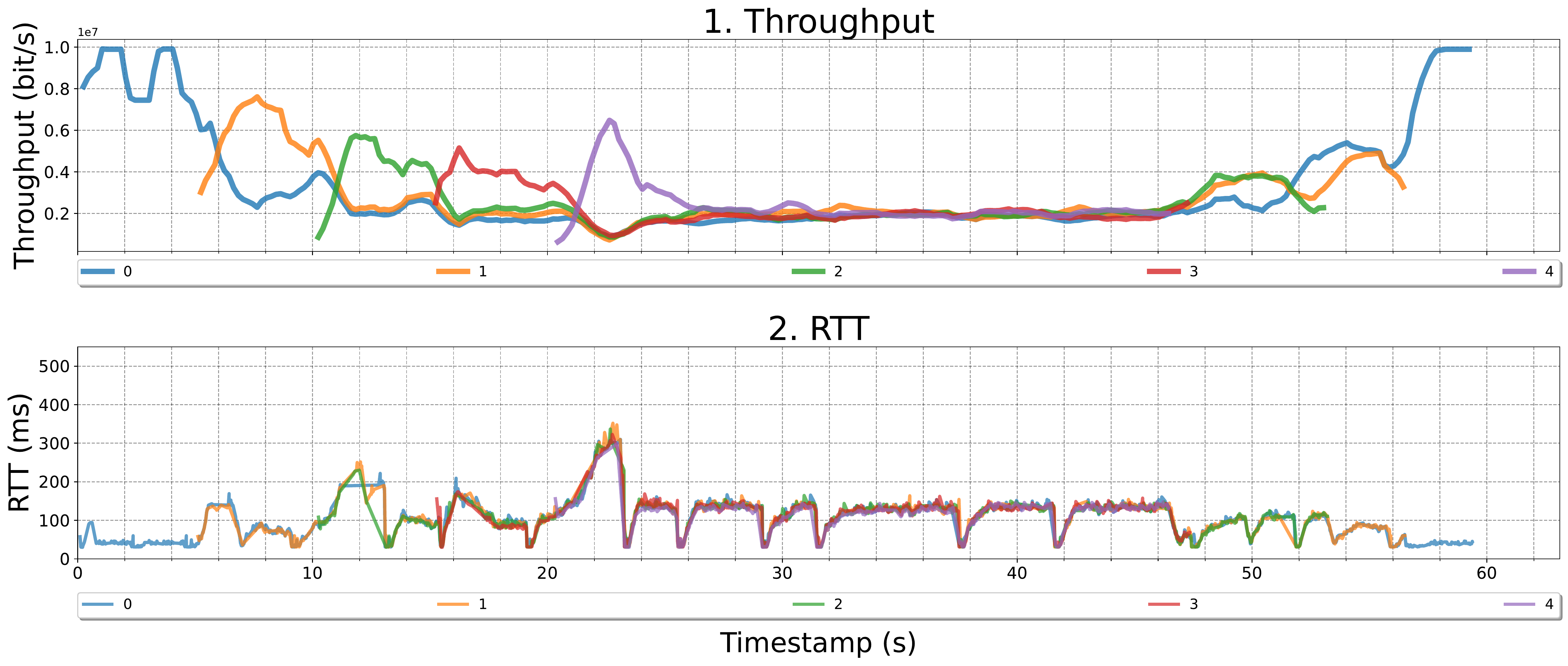}
		\caption{PPO-BBR}
	\end{subfigure}
	\caption{Multiple flows joining and leaving. Convergence of throughput and RTT for (a) the vanilla BBR and (b) PPO-BBR.}
\label{fig:multiflow}
\end{figure*}

{\noindent\bf Comparison of PPO-BBR with Vanilla BBR in Multiple Flows.} We evaluate our prototype when multiple flows join and leave in sequence. In this experiment, we focus on PPO-BBR which mostly outperform other RL algorithms in the previous experiment. Each flow in our PPO-BBR experiment is deployed with the same pre-trained agent. Figure~\ref{fig:multiflow} shows the throughput and RTT when flows join every $5$ seconds at the beginning of the emulation, and leave one-by-one every $5$ seconds at the end. The throughput of PPO-BBR converges much faster than the vanilla BBR when new flows join, and more significantly, when flows leave, it can more quickly detect the availability in bandwidth. For example, when the fifth flow joins, the flows converge to their fair share in around $5$ seconds for PPO-BBR and in around $13.5$ seconds ($2.7$x) for the vanilla BBR. Furthermore, the RTT plots show that PPO-BBR maintains a significantly lower RTT throughout the experiment. It reduces the peak RTT by $40\%$, and reacts to peak RTT $10$x faster than vanilla BBR.

\section{Conclusion}
We presented an intelligent system for automatically optimizing the parameters of congestion control protocols. Using the PPO algorithm, it dynamically improves the performance based on the state of the network and facilitates the use of congestion control protocols for new environments. In our case study, tuning BBR enhances the RTT estimation and throughput drops in a network with frequent changes. Also, the flows converge to their fair shares faster while maintaining significantly lower RTTs. We leave tuning other parameters as well as other congestion control protocols as future work.
\bibliographystyle{IEEEtran}
\bibliography{ref}

\end{document}